\documentstyle[aps,preprint,epsfig]{revtex}
\def\nt{\hbox{$\nu_\tau$ }}
\def\VEV#1{\left\langle #1\right\rangle}

\def\ie{\hbox{\it i.e., }} 
\def\eq#1{{Eq.\ (\ref{#1})}}
\def\fig#1{{Fig.\ \ref{#1}}}

\def\nps#1#2#3{        {\it Nucl. Phys. B (Proc. Suppl.) }{\bf #1} (19#2) #3}
\def\np#1#2#3{           {\it Nucl. Phys. }{\bf #1} (19#2) #3}
\def\pl#1#2#3{           {\it Phys. Lett. }{\bf #1} (19#2) #3}
\def\pr#1#2#3{           {\it Phys. Rev. }{\bf #1} (19#2) #3}
\def\prep#1#2#3{         {\it Phys. Rep. }{\bf #1} (19#2) #3}
\def\prl#1#2#3{          {\it Phys. Rev. Lett. }{\bf #1} (19#2) #3}

\def\Zp#1#2#3{           {\it Z. Physik }{\bf #1} (19#2) #3}

\def\ppnp#1#2#3{           {\it Prog. Part. Nucl. Phys. }{\bf #1} (19#2) #3}

\relax

\begin{document}
\newcommand{\ptmis}{{ {\rm p} \hspace{-0.525 em} \raisebox{-0.27 ex} {/}_{\text{\small T}} }}
\preprint{
\font\fortssbx=cmssbx10 scaled \magstep2
\hbox to \hsize{
\hfill$\vcenter{
                \hbox{\bf hep-ph/9710545}
                \hbox{\bf IFUSP 1278}
                \hbox{\bf FTUV/97-49}
                \hbox{\bf IFIC/97-50}
                }$ }}

\title{\vspace*{0cm}
$R$--Parity Violating Signals for Chargino Production at LEP II}

\author{
F. de Campos$^a$\footnote{E-mail fernando@feg.unesp.br},
O. J. P. \'Eboli$^b$\footnote{\small E-mail eboli@fma.if.usp.br},
M. A. Garc\'{\i}a-Jare\~no$^c$\footnote{\small E-mail 
miguel@flamenco.ific.uv.es}, 
and 
J. W. F. Valle$^c$ \footnote{\small E-mail valle@flamenco.ific.uv.es}
}

\address{
{\it $^a$Universidade Estadual Paulista, Campus de Guaratinguet\'a,  \\
Av. Dr. Ariberto Pereira da Cunha, 333, \\
12500-000, Guaratinguet\'a, S.P., Brazil}
\vskip 0.1cm
{\it $^b$ Instituto de F\'{\i}sica, Universidade de S\~ao Paulo,\\
C.P.\ 66.318, CEP 05315-970 S\~ao Paulo, Brazil.}
\vskip 0.1cm
{\it $^c$ Instituto de F\'{\i}sica Corpuscular - C.S.I.C.\\
Departament de F\'{\i}sica Te\`orica, Universitat de Val\`encia\\
46100 Burjassot, Val\`encia, Spain}
}

\maketitle
\thispagestyle{empty}

\vskip -.5cm

\begin{abstract}
\baselineskip=.3cm 
\vskip -1.cm

We study chargino pair production at LEP II in supersymmetric models
with spontaneously broken $R$--parity. We perform signal and
background analyses, showing that a large region of the parameter
space of these models can be probed through chargino searches at LEP
II. In particular, we determine the attainable limits on the chargino
mass as a function of the magnitude of the effective bilinear
$R$--parity violation parameter $\epsilon$, demonstrating that LEP II
is able to unravel the existence of charginos with masses almost up to
its kinematical limit even in the case of $R$--parity violation.  This
requires the study of several final state topologies since the usual
MSSM chargino signature is recovered as $\epsilon \to 0$.  Moreover,
for sufficiently large $\epsilon$ values, for which the chargino decay
mode $\chi^\pm \to \tau^\pm J$ dominates, we find through a dedicated
Monte Carlo analysis that the $\chi^\pm$ mass bounds are again very
close to the kinematic limit.  Our results establish the robustness of
the chargino mass limit, in the sense that it is basically
model-independent. They also show that LEP II can establish the
existence of spontaneous $R$--parity violation in a large region of
parameter space should charginos be produced.

\end{abstract}

\newpage


\section{Introduction}
 
In the {\em Minimal Supersymmetric Standard Model} (MSSM) the
conservation of a discrete symmetry called $R$--parity is imposed
\cite{mssm}. $R$--parity is related to the particle spin (S), lepton
number (L), and baryon number (B) through $R=(-1)^{(3B+L+2S)}$, being
all the standard model particles $R$--even while their superpartners
are $R$--odd. From this it follows that supersymmetric particles must
be produced only in pairs, with the lightest one being stable. So far,
most searches for supersymmetric particles have assumed conservation
of $R$--parity, however, neither gauge invariance nor supersymmetry
(SUSY) require its conservation. In general, we can build models
exhibiting $R$--parity violation which may be explicit \cite{expl} or
spontaneous \cite{aul}, or even the residual effect of a more
fundamental unified theory \cite{expl0}.

One possible scenario for spontaneous $R$--parity breaking is that it
takes place through nonzero vacuum expectation values (VEVs) of scalar
neutrinos \cite{beyond}.  In this case there are two distinct
possibilities depending whether lepton number is a gauge symmetry or
not.  If lepton number is part of the gauge symmetry there is an
additional gauge boson which acquires mass via the Higgs
mechanism. Therefore, there is no physical Goldstone boson and the
scale of $R$--parity violation, in the TeV range, also characterizes
the new gauge interaction \cite{ZR,RPCHI}. In this work, we consider
the alternative scenario where spontaneous $R$--parity violation
occurs in the absence of an additional gauge symmetry, so that there
is a physical massless Nambu-Goldstone boson, called Majoron
\cite{MASIpot3,MASI,ROMA,mono,NPBTAU,RPMSW} \footnote{ There are
many models where neutrinos get mass from spontaneous breaking of
lepton number \cite{fae}. In the present context the Majoron appears
also because the (tau) neutrino mass arises as a result of the
spontaneous violation of lepton number implied by the nonzero
sneutrino VEVs.}.  In this model, the Majoron remains massless and
stable in the absence of further explicit $R$--parity violating terms
that might arise, for instance from gravitational effects
\cite{rpgrav,kev}. Thus, it will lead to a missing energy signal at 
high energy accelerators.

In Majoron models, the neutralino is unstable and for moderate
strengths of the $R$--parity violating interactions, it will decay
inside the detector, either via
\begin{eqnarray}
\chi^0 &\to 
\nu_\tau Z^\star &\to  \nu_\tau \nu \nu \hbox{ , } \nu_\tau
\ell^+ \ell^- \hbox{ , } \nu_\tau q \bar{q} \;\; ;
\\
\chi^0 &\to  \tau W^\star &\to  \tau \nu_i \ell_i \hbox{
  , } \tau q \bar{q}^\prime \;\; ;
\nonumber  
\end{eqnarray}
or through Majoron emission
\begin{equation}
\label{inv}
\chi^0 \to  \nu  J \;\; .
\end{equation}
Note this decay mode is R--parity conserving, since the Majoron is
mainly R-odd (see \eq{maj}).

In the first case the neutralino gives rise to visible signals, except
for the 3 $\nu$ decay mode. In the case of \eq{inv} the neutralino
decay leads to a missing energy signature, exactly as the stable MSSM
neutralino.

In this work, we study the implications of $R$--parity breaking SUSY
models with a Majoron for chargino searches at LEP II.  In this
case, in addition to the conventional MSSM chargino decay mode
\begin{equation}
\chi^+ \to  W^+ \chi^0 \;\; ,
\label{Wchi0}
\end{equation} 
where the $W$ can be real or virtual depending on the chargino and
neutralino masses, there is a new R--parity conserving two-body decay
mode
\begin{equation}
\label{twob}
\chi^\pm \to  \tau^\pm J \;\; ,
\label{2body}
\end{equation}
As in \eq{inv} the decay in \eq{2body} is R--parity conserving it can
therefore be quite sizeable.

Note that $R$--parity violating decays $\chi^+ \to W^+ \nu_\tau$ and
$\chi^\pm \to Z \tau^\pm$ are typically negligible compared to the
above decay modes, as shown explicitly in ref.~\cite{gluino paper}.

We evaluate the LEP II potential for probing the $R$--parity violating
SUSY parameter space through the study of new signatures arising from
chargino pair production and its corresponding cascade decay. We
determine the limits on the chargino mass ($m_{\chi^+}$) for different
values of the $R$--parity violating interactions. In our analyses, we
recover the MSSM chargino mass limit, which is close to the kinematic
limit, for sufficiently small strengths of the $R$--parity violating
interactions.  As the magnitude of $R$--parity violation becomes
larger, new final state topologies become available. By performing
a Monte Carlo analysis we show that these new topologies also lead to
bounds on the chargino mass close to the kinematical limit. Assuming
unification of the gaugino mass parameters we also determine the
corresponding neutralino mass limit.


\section{Basic Framework}

We adopted the conceptually simplest model for the spontaneous
violation of $R$ proposed in Ref.\ \cite{MASIpot3} in which, by
construction, neutrinos are massless before breaking of $R$--parity.
As a result all $R$--parity violating observables are directly
correlated to the mass of the tau neutrino with the magnitude of this
correlation depending upon the choice of the $R$--parity SUSY
parameters.  Apart from the theoretical attractive of giving a
dynamical origin for the violation of $R$--parity and neutrino mass,
these models offer the possibility of realizing a radiative scenario
for the breaking of $R$--parity, similar to that of electroweak
breaking \cite{ara}.

In order to set up our notation, we first recall some basic
ingredients. The superpotential, which conserves $total$ lepton number
and $R$, is given by 
\begin{equation} 
h_u Q H_u u^c + h_d H_d Q d^c + h_e \ell H_d e^c + (h_0 H_u H_d - 
{\epsilon^\prime}^2)
\Phi + h_{\nu} \ell H_u \nu^c + h \Phi S \nu^c + \mbox{h.c.} 
 \;\; ,
\label{P}
\end{equation} 
where the couplings $h_u,h_d,h_e,h_{\nu},h$ are arbitrary matrices in
generation space.  The additional chiral superfields $(\Phi$,
${\nu^c}_i$, $S_i)$ are singlets under $SU(2) \otimes U(1)$ and carry
a conserved lepton number assigned as $(0,-1,1)$ respectively.  Note
that terms such as $\Phi^2$ and $\Phi^3$ are in principle allowed and
have been discussed in earlier papers. For example a $\Phi^2$ term was
included in ref.~\cite{RPMSW} and a $\Phi^3$ has been included in a
recent formulation of the theory with radiative symmetry breaking
\cite{ara}. However the presence of the $\Phi$ field  is not 
essential in the formulation of the theory. In schemes with radiative
breaking one may simply add bare mass terms $\mu H_u H_d$ and $M S
\nu^c$ without adding the $\Phi$ field. From the point of view
of our analysis the presence of $\Phi$ has basically no effect, as it
relies mainly on the chargino sector.

The superfields $\nu^c, S$ \cite{SST} and $\Phi$ \cite{BFS} are
required to drive the spontaneous violation of $R$--parity in an
acceptable way, so that the Majoron is mostly a singlet, that is given
by the imaginary part of \cite{MASIpot3}
\begin{equation}
\frac{v_L^2}{Vv^2} (v_u H_u - v_d H_d) +
              \frac{v_L}{V} \tilde{\nu_{\tau}} -
              \frac{v_R}{V} \tilde{\nu^c}_{\tau} +
              \frac{v_S}{V} \tilde{S_{\tau}} \;\; ,
\label{maj}
\end{equation}
where the isosinglet VEVs
\begin{equation}
\begin{array}{lr}
v_R = \VEV {\tilde{\nu}_{R\tau}}\:, &
v_S = \VEV {\tilde{S_{\tau}}} \;\; ,
\end{array} 
\end{equation}
and $V = \sqrt{v_R^2 + v_S^2}$ characterizes $R$ or lepton number
breaking. The isodoublet VEVs
\begin{equation}
\begin{array}{lr}
v_u = \VEV {H_u} \:, &
v_d = \VEV {H_d} 
\end{array} 
\end{equation}
are responsible for the breaking of the electroweak symmetry and the
generation of fermion masses with the combination $v^2 = v_u^2 +
v_d^2$ being fixed by the $W,Z$ masses. Finally, there is a small seed
of $R$--parity breaking in the doublet sector, \ie
\begin{equation}
v_L = \VEV {\tilde{\nu}_{L\tau}} 
\end{equation}
whose magnitude is now related to the Yukawa coupling $h_{\nu}$. Since
this vanishes as $h_{\nu} \to  0$, we can naturally satisfy the
limits originating from stellar energy loss \cite{KIM}.  Note that,
unlike the standard seesaw model, the neutral leptons members of the
singlet superfields ${\nu^c}_i$ and $S_i$ are given only Dirac-type
masses.

Notice that we have assumed R--parity violating VEVs only for the
third generation. This is the theoretically well-motivated choice if
one has in mind a radiatively induced symmetry breaking mechanism
\cite{ara,bilrad}, since the largest Yukawa couplings are those 
of the third generation
\footnote{ Some of the effects in such a complete dynamical scheme 
get communicated through mixing to the lightest generations. See, for
example, ref.~\cite{NPBTAU}.}. For future use we define an effective
parameter $\epsilon_i \equiv h_{\nu ij} v_{Rj}$, which measures the
violation of $R$--parity now expressed as an effective bilinear
superpotential term which breaks R--parity explicitly. Together with
the standard MSSM $\mu$ parameter it will affect the fermion mass
matrices given below.

The form of the chargino mass matrix is common to a wide class of
$SU(2) \otimes U(1)$ SUSY models with spontaneously broken $R$--parity
and is given by
\begin{equation}
\begin{array}{c|cccccccc}
& e^+_j & \tilde{H^+_u} & -i \tilde{W^+}\\
\hline
e_i & h_{e ij} v_d & - h_{\nu ij} v_{Rj} & \sqrt{2} g_2 v_{Li} \\
\tilde{H^-_d} & - h_{e ij} v_{Li} & \mu & \sqrt{2} g_2 v_d\\
-i \tilde{W^-} & 0 & \sqrt{2} g_2 v_u & M_2
\end{array}
\;\;\;\;\;\; .
\label{chino}
\end{equation}
Two matrices U and V are needed to diagonalise the $5 \times 5$ 
(non-symmetric) chargino mass matrix
\begin{eqnarray}
{\chi}_i^+ = V_{ij} {\psi}_j^+ \;\; ,
\\
{\chi}_i^- = U_{ij} {\psi}_j^- \;\; ,
\label{INO}
\end{eqnarray}
where the indices $i$ and $j$ run from $1$ to $5$, $\psi_j^+ = (e_1^+,
e_2^+ , e_3^+ ,\tilde{H^+_u}, -i \tilde{W^+}$) and $\psi_j^- = (e_1^-,
e_2^- , e_3^-, \tilde{H^-_d}, -i \tilde{W^-}$).

If the  singlet superfield mass terms are large one can truncate
the neutralino mass matrix so as to obtain an effective $7\times 7$
matrix of the following form \cite{MASIpot3}
\begin{equation}
\begin{array}{c|cccccccc}
& {\nu}_i & \tilde{H}_u & \tilde{H}_d & -i \tilde{W}_3 & -i \tilde{B}\\
\hline
{\nu}_i & 0 & h_{\nu ij} v_{Rj} & 0 & g_2 v_{Li} & -g_1 v_{Li}\\
\tilde{H}_u & h_{\nu ij} v_{Rj} & 0 & - \mu & -g_2 v_u & g_1 v_u\\
\tilde{H}_d & 0 & - \mu & 0 & g_2 v_d & -g_1 v_d\\
-i \tilde{W}_3 & g_2 v_{Li} & -g_2 v_u & g_2 v_d & M_2 & 0\\
-i \tilde{B} & -g_1 v_{Li} & g_1 v_u & -g_1 v_d & 0 & M_1
\end{array}
\;\;\;\;\;\; 
\label{nino}
\end{equation}
where $M_{1(2)}$ denote the supersymmetry breaking gaugino mass
parameters and $g_{1(2)}$ are the $SU(2) \otimes U(1)$ gauge couplings
divided by $\sqrt{2}$. Moreover, we assumed the canonical GUT relation
$M_1/M_2 = \frac{5}{3} \tan^2{\theta_W}$. We have however included the
full neutral mass matrix, including the singlet sector, and
diagonalized it numerically in order to determine, for example, the
\nt mass and to identify the physical mass eigenstates. In any case
the singlets are hardly relevant for our present analysis, as they
appear in the chargino mass matrix only through the effective bilinear
parameters$\epsilon_i \equiv h_{\nu ij} v_{Rj}$, which measures the
violation of $R$--parity and the usual MSSM $\mu$ parameter.

The matrix (\ref{nino}) is diagonalised by a $7 \times 7$ unitary
matrix N,
\begin{equation}
{\chi}_i^o = N_{ij} {\psi}_j^o \;\; ,
\end{equation}
where $\psi_j^0 = ({\nu}_i,\tilde{H}_u,\tilde{H}_d,-i \tilde{W}_3,-i
\tilde{B}$), with $\nu_i$ denoting the three weak-eigenstate
neutrinos.

In our analyses, we considered typical values for the SUSY parameters $\mu$,
$M_2$ that can be covered by chargino production at LEP II:
\begin{equation}
\label{param21}
\begin{array}{cccc}
-200 &\leq~ \mu  &\leq 200 &\mbox{ [GeV]} \;\; , \\
40   &\leq~ M_2  &\leq 400 &\mbox{ [GeV]} \;\; . \\
\end{array}
\end{equation}
We also varied $\tan \beta$ in the range
\begin{equation}
\label{beta}
2 \leq~ \tan\beta=\frac{v_u}{v_d} \leq 40 \;\; .
\end{equation}
This is a standard choice for the ranges of the SUSY parameters which
generously accounts for chargino masses within the kinematical reach
of LEP I. This range has only been used in order to have an overview
of parameter space in the first two figures of our paper (see
below). Note that we have explicitly limited $\tan\beta$ to values
that are consistent with supergravity versions with perturbative
Yukawa couplings up to the GUT scale, excluding, for example
$\tan\beta=1$. No essential change would result if lower $\tan\beta$
values were included. In the analysis of the signals we have simply
fixed $\tan \beta$ at the two illustrative values used by the DELPHI
collaboration.

As we can see from the neutralino and chargino mass matrices, the
$\epsilon$ parameter gives the main contribution to the mixing between
charged (neutral) leptons and the charginos (neutralinos) and also
leads to $R$ violating gauge couplings.

We have required the parameters $h_{\nu i,3}$ and the expectation
values lie in the ranges
\begin{equation}
\label{param2}
\begin{array}{cccc}
10^{-10}\leq h_{\nu 13},h_{\nu 23} \leq 10^{-1} & & & 
10^{-5}\leq h_{\nu 33} \leq 10^{-1}\\
\end{array}
\end{equation}
\vskip -24pt
\begin{equation}
\label{param1}
\begin{array}{cccc}
v_L=v_{L3}=100\:\: \mbox{MeV} 
&&& \\
50\:\: \mbox{GeV}\leq v_R=v_{R3}\leq 1000 \:\: \mbox{GeV} \ 
&&& \\
50\:\: \mbox{GeV}\leq v_S=v_{S3}=v_R\leq 1000 \:\: \mbox{GeV} 
\end{array}
\end{equation} 
For definiteness we have set $v_{L1}=v_{L2}=0$ and $v_{R1}=v_{R2}=0$.

The above range for the $R$--parity breaking parameters is quite
reasonable and generous, and has been used widely in previous papers
e.g. \cite{RPCHI,RPMSW}. There are many restrictions on the parameters
in broken $R$ models which follow from laboratory experiments related
to neutrino physics, weak interactions, cosmology, and astrophysics
\cite{beyond,fae}.  The most relevant constraints come from
neutrino-less double beta decay and neutrino oscillation searches,
direct searches for anomalous peaks at $\pi$ and K meson decays, the
limit on the tau neutrino mass \cite{eps95}, and cosmological limits
on the $\nu_\tau$ lifetime and mass, as well as limits on muon and tau
lifetimes, on lepton flavour violating decays, and universality
violation. These constraints have been taken into account in several
previous papers \cite{NPBTAU,RPMSW}. Here we have just included an
updated version.

The model described above constitutes a very useful way to parametrise
the physics of $R$ violation, due to the strict correlation between
the magnitude of $R$ violating phenomena and the resulting $\nu_\tau$
mass.  In other words, neutrinos are strictly massless before breaking
$R$, therefore all $R$ violating observables, such as the lightest
neutralino decay rate $\Gamma_\chi$, are directly correlated to the
mass of the tau neutrino. In fact, the $\tau$ neutrino mass may be
written schematically as $m_{\nu_\tau} \sim \xi
\epsilon^2/{m_{\chi^+}}$, where $\xi$ is some effective parameter
given as a function of $M_2$, $\mu$, $\tan\beta$, etc \footnote{For a
more complete discussion see the second paper in Ref.\
\cite{ara,bilrad}}. This establishes a correlation between the violation of
$R$ and the $\nu_\tau$ mass showing explicitly how the broken
$R$-model provides an interesting mechanism to understand the origin
of neutrino mass without invoking physics at very high energy scales
\cite{fgjrv}.

\begin{figure}
\begin{center}
\mbox{\epsfig{file=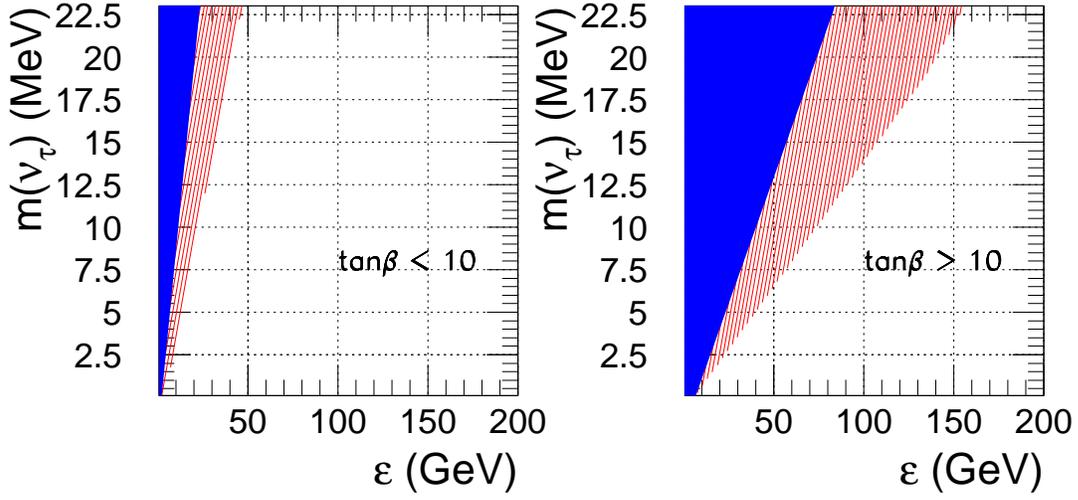,width=\linewidth}}
\end{center}
\caption{The value of the predicted tau neutrino mass in our model
is compatible with the LEP experimental limits in the light shaded
area.  Within the dark shaded area, chargino masses are such that they
can be produced at $ \protect\sqrt{s}=172 $ GeV. }
\label{fig:mntau_eps}
\end{figure}

In Fig.\ \ref{fig:mntau_eps}, we exhibit the tau neutrino mass as a
function of $\epsilon$, showing in light grey the region in the
($m_{\nu_{\tau}}$,$\epsilon$) plane which is compatible with the tau
neutrino mass limit from LEP. We also present in this figure the
region in which the charginos can be pair produced at LEP, which
corresponds to a smaller range of $\epsilon$ values (dark zone).  As
we can see, for $\tan\beta < 10$, the maximum value of $\epsilon$ that
can be probed through chargino pair production at LEP II is around
$20$ GeV and it increases for larger $\tan\beta$.  For definiteness we
fixed the value of $\epsilon$ in our analysis.

In the following section we describe the most relevant chargino and
neutralino decay modes for this work.  A complete list of the decay
widths and couplings can be found in \cite{gluino paper} or
\cite{lepsensi}.


\section{Signals and Backgrounds}

\subsection{Chargino Production}

\begin{figure}
\begin{center}
        \mbox{\epsfig{file=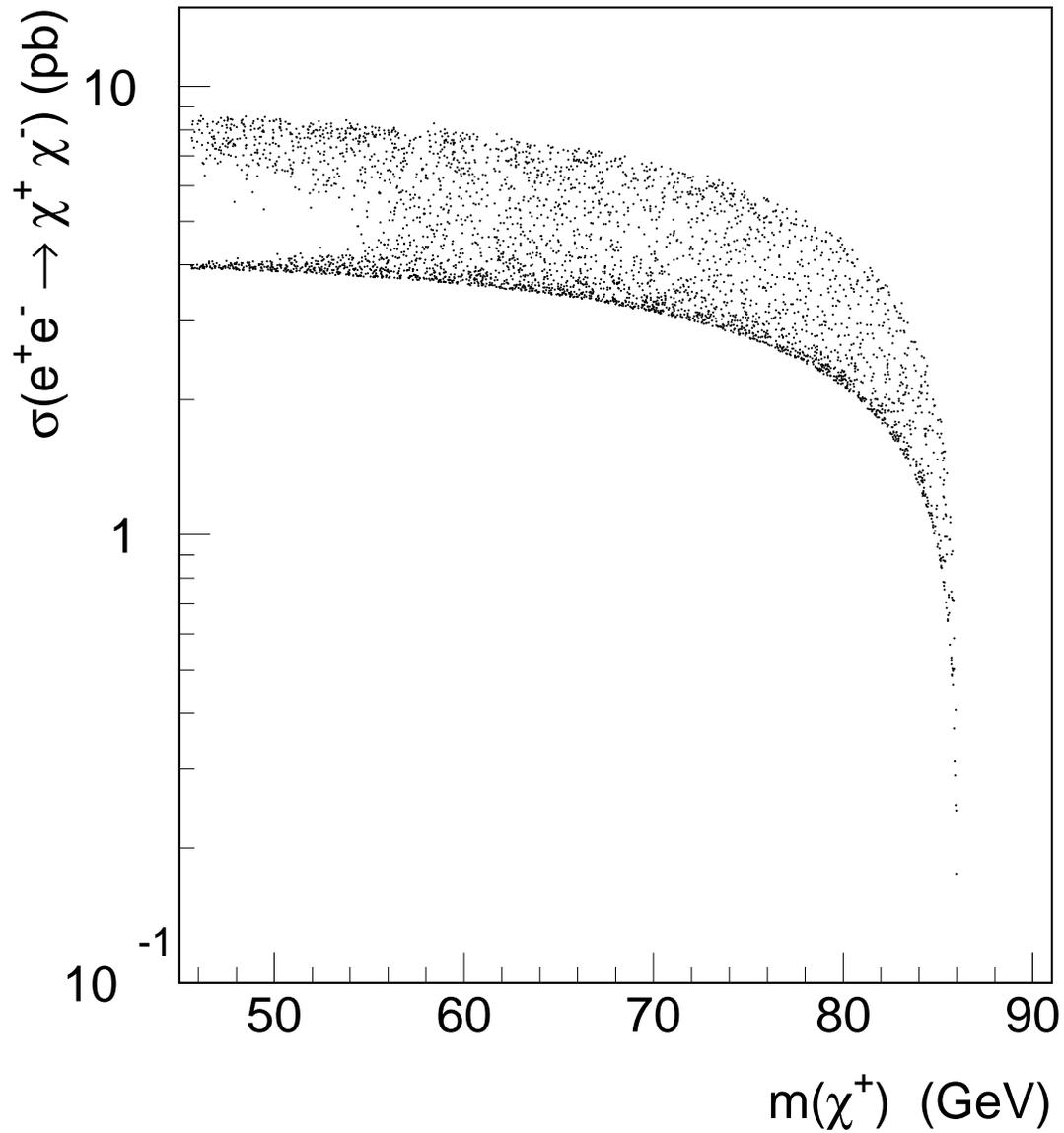,width=\linewidth}}
\end{center}
\caption{$e^+ e^-  \to  \chi^+ \chi^- $ 
  cross section, in the large $m_{\tilde{\nu}}$ limit, versus chargino
  mass for the parameter region defined in Eqs.\
  (\protect\ref{param2}) and (\protect\ref{param1}) and
  $\protect\sqrt{s}=172$ GeV.  The upper and lower limiting curves of
  this plot define the range of LEPII chargino pair production cross
  section for our parameter space.}
\label{fig:xs}
\end{figure}

At LEP II the lightest chargino may be pair produced via
\begin{equation}
e^+ e^- \to  \gamma,Z,\tilde{\nu} \to  \chi^+ \chi^-
\;\; .
\end{equation}
In this work we assumed that the sneutrinos are so heavy that only the
$\gamma$ and $Z$ s-channels contribute to the cross section; see, for
instance, Refs.\ \cite{bartl} and \cite{Navas}. In fact, the contribution of
the t-channel to the total cross section is completely negligible for
sneutrinos heavier than 500 GeV.  Fig.\ \ref{fig:xs} shows a scatter plot of
the allowed values of the $e^+ e^- \to  \chi^+ \chi^- $ total cross
section versus the chargino mass for $\sqrt{s}=172$ GeV, when the parameters
are varied as in \eq{param21} and \eq{beta}. As one can see, this cross
section varies between 2 and 10 pb for almost all kinematically allowed
chargino masses.

Although, our model allows the  single $R$--parity-violating chargino
production
\begin{equation}
e^+ e^- \to  \chi^{\pm} \tau^{\mp} \;\; ,
\end{equation}
we only considered in this paper the pair-production of charginos, as
in the MSSM, since the cross section for the single chargino
production at LEP II is too small to be observed \cite{adv}.


\subsection{Neutralino and Chargino Decays}

The breaking of $R$--parity not only opens new decay channels for the
chargino but also allows the lightest neutralino to decay. Therefore,
there are new signatures for SUSY, some of them being very striking.
In order to simplify the analysis, we assumed that all sfermions are
sufficiently heavy not to influence the physics at LEP II, {\em i.e.}
we neglected their effects in the chargino production as well as in
their decays.  In the present model, the lightest neutralino ($\chi^0$)
can decay invisibly $\chi^0 \to  \nu J$, as in \eq{inv}, as
well as into $R$--parity violating channels
\begin{eqnarray}
\chi^0 &\to  \nu_\tau Z^\star  
&\to   \nu_\tau \nu \nu \hbox{ , } \nu_\tau \ell^+ \ell^-
\hbox{ , } \nu_\tau q \bar{q} \;\; ;
\\
\chi^0 &\to   \tau W^\star
&\to   \tau \nu_i \ell_i \hbox{ , } \tau q \bar{q}^\prime
\;\; .
\end{eqnarray}
For the chargino masses accessible at LEP II, the above $W$ and $Z$ are
off-shell and the neutralino has only two-body majoron decays and the above
three-body modes. A complete description of neutralino decay modes as a
function of the model parameters and masses can be found in Ref.\
\cite{gluino paper}.

It is interesting to notice that all three-body decay channels of the
lightest neutralino are {\em visible}, except for the neutral current
one leading to 3 neutrinos. In the parameter space regions where most
of neutralino decays are {\em visible}, the strategies to search for
SUSY particles are considerably modified with respect to ones used in
the MSSM. The MSSM is recovered as a special limit of this class of
models, when the lightest neutralino decays outside the detector
because $R$--parity violation is not strong enough.  Notwithstanding,
$\chi^0$ decays can also lead to missing momentum due to the presence
of neutrinos or Majorons.  It is important to notice that, the
neutralino of a spontaneously broken $R$--parity model fakes the MSSM
one when the invisible decay given in Eq.\ (\ref{inv}) dominates since
its decay products escape undetected.


In $R$--parity breaking models, the decays of the lightest chargino, denoted
by $\chi^\pm$, are modified by the existence of new channels.  In models with
a majoron, the lightest chargino ($\chi^\pm$) exhibits the two-body decay mode
$\chi^\pm \to  \tau^\pm J$ of \eq{twob}, in addition to the
channels\footnote{Notice that there is the possibility of a chargino decaying
into the second lightest neutralino plus a $W^{+}$, which conserves
$R$--parity. However, for the parameter range considered, the second lightest
neutralino mass is around the chargino mass, so that this decay is forbidden
or kinematically suppressed. }
\begin{eqnarray}
\chi^\pm &\to   \nu_\tau W^\star 
&\to  \nu_\tau q \bar{q}^\prime \hbox{ , } 
\nu_\tau \ell^\pm_i \nu_i
\;\; , \\
\chi^\pm &\to   \tau^\pm Z^\star 
&\to  \tau^\pm q \bar{q} \hbox{ , } \tau^\pm \ell^+ \ell^-
\hbox{ , } \tau^\pm \nu \bar{\nu}
\;\; , \\
\chi^\pm &\to   \chi^0 W^\star 
&\to  \chi^0 q \bar{q}^\prime \hbox{ , } 
\chi^0 \ell^\pm_i \nu_i
\;\; ,
\end{eqnarray}
where we again assumed that the sfermions are heavy. In the framework
of the MSSM only the last decay channel is present, however, with the
$\chi^0$ being stable. Therefore, the breaking of $R$--parity can
modify substantially the signature of charginos.

For the sake of illustration we exhibit in Fig.\ \ref{fig:br} typical
values of the branching ratios of charginos and neutralinos, when we
vary $\epsilon$ for $\mu=150$ GeV, $M_2=100$ GeV, and $\tan\beta=35$.
For neutralinos we exhibit its total visible and invisible branching
ratios, where we included in the invisible width the contributions
coming from the neutrino plus majoron channel ($\chi^0 \to \nu$J), as
well as from the neutral current channel when the $Z$ decays into a
pair of neutrinos ($\chi^0 \to 3 \nu$).

For small $\epsilon $ values (up to $10^{-4}$ GeV) the neutralino will
decay outside the detector (since its lifetime is larger than
$10^{-6}$ s) so that it leaves no visible track. In this case it is
effectively stable and the MSSM limit is restored. In this region the
invisible component of the neutralino decay is associated only to the
$\chi^0 \to 3 \nu$ channel.  When the $\epsilon$ parameter grows up to
the order of $1$ GeV the decay channels get mixed and both neutralinos
as well as charginos have $R$--parity violating decays at the same
level as the standard MSSM ones. As expected, above $\epsilon \sim 1$
GeV or so $\chi^\pm$ and $\chi^0$ decay predominantly into majorons,
that is, the invisible channel dominates the decay of the neutralino
and the main chargino decay mode is $\tau^+ J$.


\subsection{Signatures for chargino pair production}

At LEP II, there is a variety of topologies associated to the production of
lightest chargino pairs. We classified the possible signatures into four
categories which contain almost all the final states allowed in $R$--parity
violating models.

\begin{itemize}

\item {\em MSSM topologies:} This class includes the following topologies
\begin{eqnarray*}
\chi^+ \chi^- &\to & 4\hbox{ jets } + \ptmis \;\; ,
\\ 
\chi^+ \chi^- &\to & 2\hbox{ jets } + \ell^\pm + \ptmis \;\; ,
\\ 
\chi^+ \chi^- &\to & \ell^+ \ell^- + \ptmis \;\; ,
\end{eqnarray*}
where $\ell^\pm$ stands for $e^\pm$ or $\mu^\pm$. These are the channels used
in the chargino searches within the framework of the MSSM. In majoron models,
such topologies are obtained by charged-current decays into $\nu_\tau W^*$ or
$\chi^0 W^*$, with the $\chi^0$ decaying invisibly. As we can see from Fig.\
\ref{fig:br}, these topologies are expected to be important for very small
values of $\epsilon$ since this is the region where $\chi^\pm$ decays
predominantly into $W^* \chi^0$ and the neutralino has such a long life-time
and it is not observed in the detector. These topologies also play an
important r\^ole for moderate values of $\epsilon$ (e.g.  $\simeq 0.1$) where
the invisible decay of the neutralino is dominant and the chargino still
decays into a $\chi^0 W^*$.


\item {\em Multi-fermion (exotic) topologies:} When the neutralino
  decays visibly, almost all the three-body channels of the chargino lead to
  at least 3 charged leptons or jets. Again, this occurs for small values of
  $\epsilon$, where the chargino decays predominantly into $\chi^0 W^*$.
  Therefore, the pair production of charginos can give rise to events with a
  large multiplicity of leptons and/or jets in this region of the parameter
  space. This is a striking signature of new physics.  We focussed our
  attention on final states with 5 or more charged leptons and/or jets that
  also present missing energy.

  
\item {$\tau^\pm$ with 2 jets topology:} For moderate values of
  $\epsilon$ the neutralino decays invisibly and the chargino either
  into $\tau^\pm J$ or into $\chi^0 W^*$. An important topology to
  analyze for this range of parameters is
\begin{equation}
\chi^+ \chi^- \to  \tau^\pm + 2 \hbox{ jets } + \ptmis \;\; ,
\end{equation}
which arises when one of the charginos decays to $\tau^\pm J$ while
the other one decays to $\chi^0 W^*$.


\item {$\tau^+ \tau^- \protect\ptmis$:} For large values of
  $\epsilon$, the chargino decay is dominated by $\chi^\pm \to 
  \tau^\pm J$, therefore, the signal arising from its pair production
  is
\begin{equation}
        \chi^+ \chi^- \to  \tau^+ \tau^- \ptmis \;\; .
\end{equation}
In this case the signal for charginos in majoron models is the same of stau
production in the MSSM framework \cite{fr}.

\end{itemize}


\subsection{Standard Model Backgrounds and Respective Cuts}
\label{SMb}

Our goal is to evaluate the potential of LEP II to unravel the existence of
supersymmetry with spontaneous $R$--parity violation. In order to do so, we
studied the signals and backgrounds, choosing the cuts to enhance the
former. The main background for the above topologies are:

\begin{itemize}
  
\item{\it MSSM topologies:} The background for these signals has been
  studied at length by several groups, including the experimental
  collaborations \cite{Navas}. The main sources of background for these
  topologies are $e^+ e^- \to  f \bar{f}~(n\gamma)$ ($f=q$ or
  $\ell^\pm$), $W^+ W^-$, $(Z/\gamma)^\star (Z/\gamma)^*$, $We\nu_e$, and
  $Ze^+e^-$. The total cross sections of the backgrounds and respective cuts
  for the three MSSM topologies, after the cuts imposed by DELPHI in their
  analysis, are given in \cite{Navas}. Moreover, we can easily obtain the
  signal cross sections in models with $R$--parity violation by evaluating the
  cross section for chargino pair production and multiplying it by the
  appropriate branching ratios and experimental detection efficiencies -- that
  is, we basically re-scale the DELPHI analysis to our scenario. These
  efficiencies are a function of the mass difference between the chargino and
  the lightest neutralino, with a small fluctuation due to the the statistical
  error in the simulation as well as an intrinsic dependence on the chargino
  mass. To be conservative, we have considered the lowest value of these
  efficiencies for each mass difference.
  
\item {\em Multi-fermion (exotic) topologies:} At the parton level,
  these events exhibit 5 or more fermions. For instance, we can have final
  states $\ell^+_i \ell^-_i q \bar{q}^\prime \ell^\pm + \ptmis$, or six
  charged leptons and missing $p_T$, or 8 jets and missing $p_T$. The Standard
  Model (SM) contributions to these events originate only from higher orders
  in perturbation theory, and consequently they have negligible cross
  sections. In our analysis, we assumed that there is no SM background and a
  conservative detection efficiency of $30$\%.
  
\item {$\tau^\pm$ plus 2 jets topology:} The SM processes that can
  give rise to this topology are $e^+ e^- \to  W^+ W^-$,
  $(Z/\gamma)^\star (Z/\gamma)^\star$, which also contribute to the $jj\ell$
  MSSM topology background. At the parton level the cross sections of these
  process are the same for $\ell^\pm=e^\pm$, $\mu^\pm$, or
  $\tau^\pm$. Therefore, we evaluated the size of this background by
  multiplying the DELPHI's result for $\sigma_{SM}(jj \mu^\pm + \ptmis)$ by a
  $\tau$ identification efficiency, which we have taken as 80\%.
  
\item {$\tau^+ \tau^- + \ptmis$:} This is the main signal of
  $R$--parity violation models over a large $\epsilon$ range. It
  happens to be the same signal which would arise from the
  pair-production of staus in the MSSM framework.  We have constructed
  an event generator to simulate the pair-production of charginos as
  well as their decays within the framework of $R$--parity violating
  models. The SM backgrounds were studied using the event generator
  PYTHIA \cite{pythia}. We considered the following SM processes,
  taking into account the QED (QCD) initial and final state radiation,
  as well as fragmentation and $\tau$ decay:
\begin{eqnarray}
e^+ e^- &\to  W^+ W^- &\to  \ell^+ \ell^- \ptmis
\; ,  \\
e^+ e^- &\to  (Z/\gamma)^\star (Z/\gamma)^\star
& \to  \ell^+ \ell^- \ptmis
\; ,  \\
e^+ e^- & \to  (Z/\gamma)^\star & \to  \ell^+ \ell^- \ptmis
\; ,  \\
e^+ e^- & \to  [e^+e^-] \gamma\gamma & \to  \ell^+ \ell^- \ptmis
\; .
\end{eqnarray}

\end{itemize}

In order to reduce these backgrounds, we applied a series of cuts similar to
the ones used by DELPHI and ALEPHI for the stau search \cite{cuts}. Initially
we kept only the events presenting ``two jets'', which might be leptons, with
a visible mass larger than 6 GeV.  We also vetoed events exhibiting photons
with more than 4 GeV whose angle with each jet is greater than 10$^\circ$ and
whose invariant mass with the jets is greater than 2 GeV.  The two-photon
background is eliminated efficiently by requiring the missing transverse
momentum to be larger than 6\% of the center-of-mass energy in events with a
visible mass smaller than 30 GeV. We also imposed that the polar angle of the
missing momentum lies between $30^\circ$ and $150^\circ$.

A very useful variable is defined by the following procedure
\cite{cuts}: first, we projected the jet momenta into the plane
perpendicular to the beam axis. Then we evaluated the thrust from the
projected momenta, and defined $\delta$ as the scalar sum of the transverse
components of the projected momenta with respect to this thrust axis. We also
defined the acoplanarity $A$ as the angle in the plane perpendicular to the
beam between the 2 jets.  With these quantities we can reduce considerably the
fermion pair background ($Z^*/\gamma^* \to \ell^+ \ell^-$) by rejecting the
events that lead to $17.1~ \delta + 120 - A < 0$. This cut eliminates a large
fraction of the fermion pair events since these tend to exhibit back-to-back
jets with a rather small $\delta$.

The $WW$ background is similar to the signal. However, we can discard a large
fraction of the $W^\pm\to \nu e^\pm$ or $\nu\mu^\pm$ events remembering that
the $e^\pm$ or $\mu^\pm$ originating from $W$'s are more energetic than the
ones coming from $\tau^\pm$ decays.  This is accomplished by requiring the
largest lepton momentum to be smaller than 22 GeV. If both $W$'s decay
leptonically, we also demanded the second lepton to have a momentum smaller
than 15 GeV.
 
After applying the above cuts and for center-of-mass energy of 172 GeV, the
$\gamma\gamma$ background is completely eliminated. On the other hand, the
cross section for fermion pair production is reduced to 13 fb, while the $ZZ$
background has a cross section of 4 fb. Most of the background events
originate from $WW$ pairs whose cross section is 60 fb. Nevertheless, the
signal possesses an efficiency of 30--40\% depending on the chargino mass.


\section{Results}

\begin{figure}

\parbox[c]{3.0in}{
\mbox{\epsfig{file=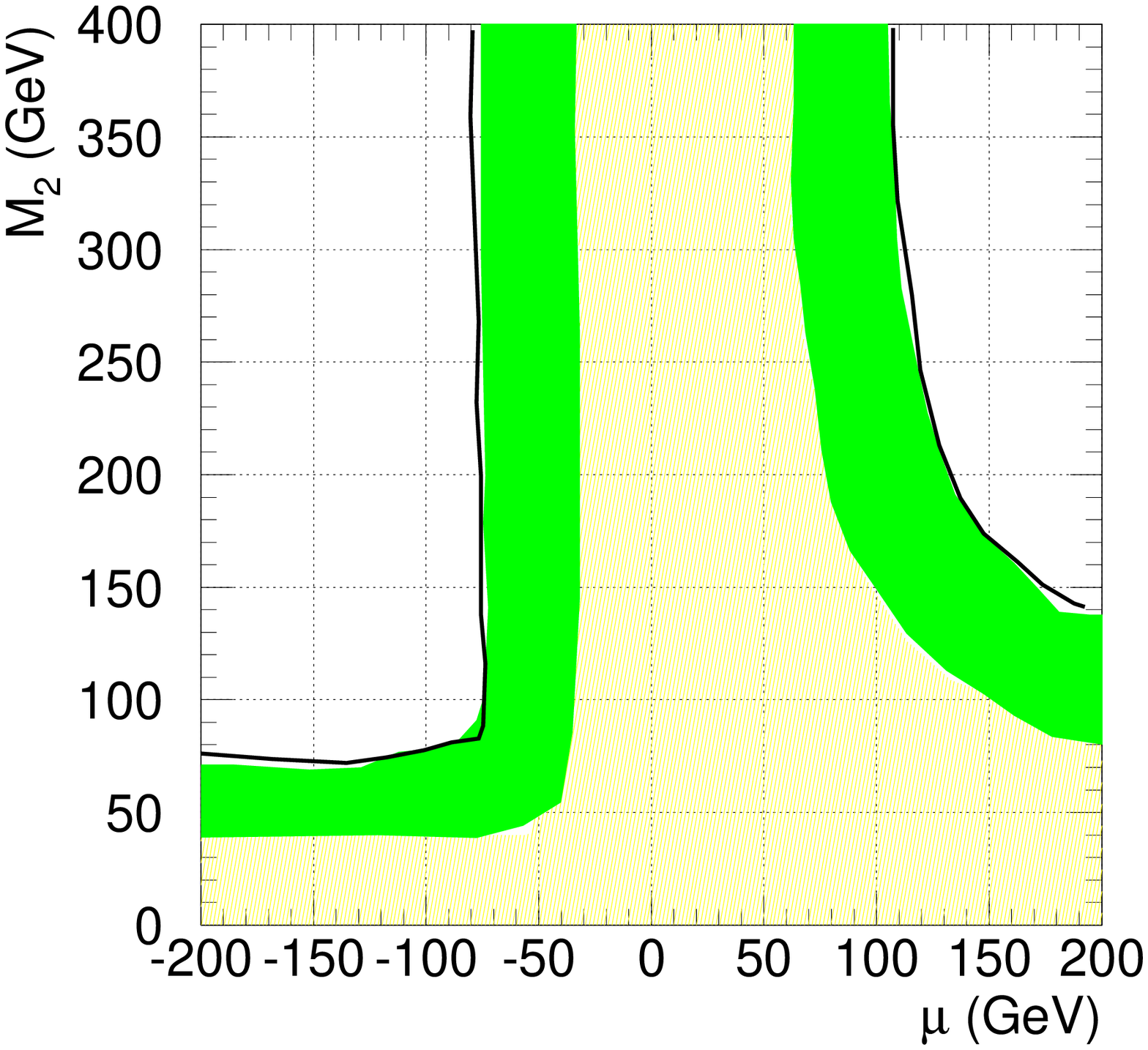,width=\linewidth}}
\begin{center}{\bf (a)}\end{center}
  }
\hfill
\parbox[c]{3.0in}{
\mbox{\epsfig{file=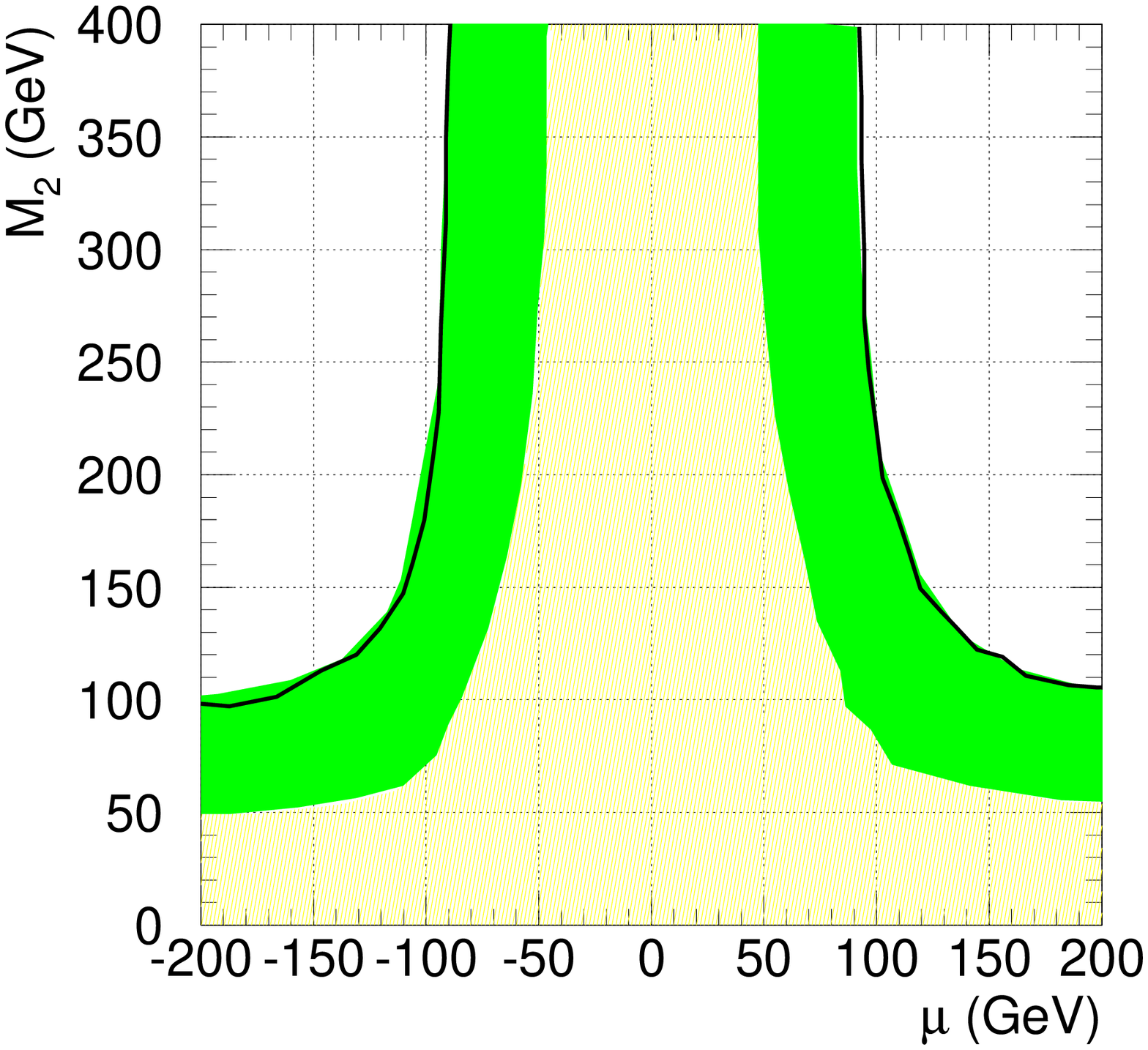,width=\linewidth}}
\begin{center}{\bf (b)}\end{center} }
\caption{95\% CL excluded region in the ($\mu $, $M_2$) plane (dark
  shaded area) in the MSSM limit $\epsilon =v_L=0$ for $\tan\beta=2$ (a)
  [$\tan\beta=35$ (b)], $\protect \sqrt{s}=172$ GeV, and an integrated
  luminosity of 300 pb$^{-1}$.  The light shaded zone is excluded in the MSSM
  limit by LEP I while the solid curve denotes the LEP II kinematical limit.}
\label{fig:zone2-0}
\end{figure}

For the sake of definiteness we considered a center-of-mass energy of 172 GeV
and a total integrated luminosity of 300 pb$^{-1}$, according to LEP II design
expectations \cite{lep2workshop}. However, our results should be a
conservative estimate of the LEP II potentiality even for actual energies and
luminosities. In analogy to the usual analyses performed for the MSSM, we
present the 95\% CL excluded regions of the $(\mu, M_2)$ SUSY parameter space
for different values of $\tan\beta$ and $\epsilon$, assuming that the number
of observed events is equal to the expected one for the background
\cite{lep2workshop}.  First of all, we obtained bounds for $\epsilon = v_R =
v_L = 0$, which should reproduce the MSSM results.  We show in Fig.\
\ref{fig:zone2-0} that we indeed obtain exactly the same limits found in the
MSSM analyses \cite{Navas}, for both $\tan\beta=2$ and 35. This shows that
we are consistent.

\begin{figure}
\begin{center}
        \mbox{\epsfig{file=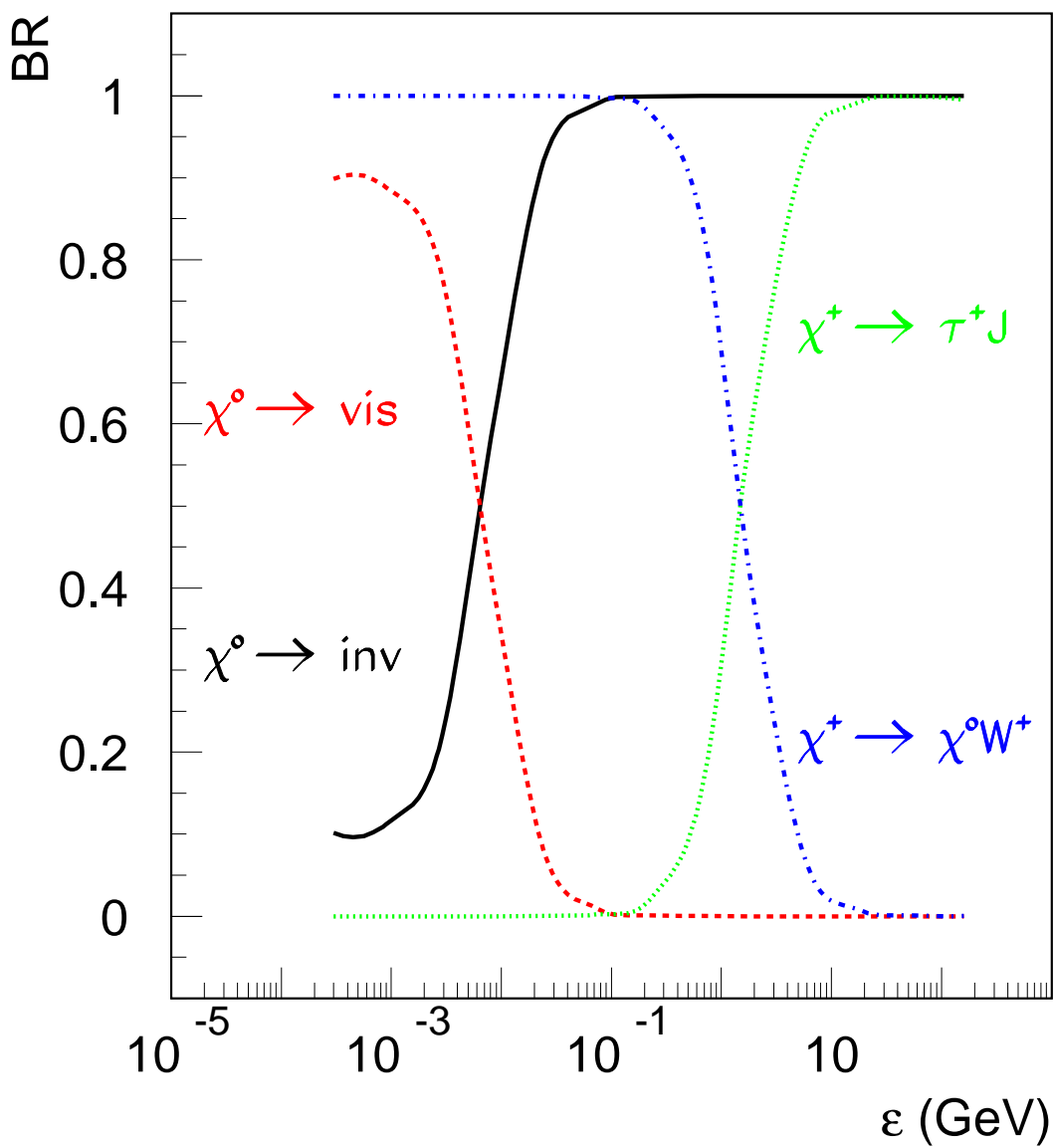,width=\linewidth}}
\end{center}
\caption{Typical neutralino and chargino decay branching ratios as a
function of $\epsilon $ for $\mu = 150$ GeV, $M_2 = 100$ GeV, and
$\tan\beta = 35$.}
\label{fig:br}
\end{figure}

For relatively small values of the $R$--parity violation parameter
$\epsilon$ the most important topologies are the MSSM and the exotic
multi-fermion ($\tau^\pm$ plus 2 jets) one for small (large) values of
$\tan\beta$. We can see from Fig.\ \ref{fig:zone2-0.1}, for $\epsilon
= 0.1$ GeV and $\tan\beta=2$, that the main constraints still come
from the MSSM final states while the exotic multi-fermion channels are
irrelevant to the final limits. This result can be understood by
looking at Fig.\ \ref{fig:br}, since for this choice of parameters the
neutralino decays mostly to $\nu J$, remaining undetected and thus
giving the conventional MSSM missing momentum signal. As $\tan\beta$
increases the importance of the multi-fermion channel diminishes while
the channel $\tau^\pm$ plus 2 jets starts to become important. We
present in Fig.\ \ref{fig:zone35-0.1} the 95\% CL excluded regions in
the plane $(\mu,M_2)$ for $\epsilon = 0.1$ GeV and $\tan\beta = 35$,
which clearly shows that the MSSM and $\tau^\pm$ plus 2 jets
topologies lead to similar bounds.

For larger values of $\epsilon$, the neutralino decays mostly
invisibly while the chargino presents a sizeable $\tau J$ branching
ratio; see Fig.\ \ref{fig:br}. Therefore, we expect that the $2\mbox{
jets} + \tau$ and $\tau \tau JJ$ signatures contribute significantly
to the chargino mass bound, while the importance of the MSSM topologies
becomes smaller.  In fact, Fig.\ \ref{fig:zone2-1} shows for $\tan
\beta= 2$ and $\epsilon=1$ GeV that the most important channel for
these parameters is $2\mbox{ jets} + \tau$ in a large fraction of the
parameter space. However, for this value of $\tan \beta$, the MSSM
final states still lead to important bounds for small values of $M_2$.
Moreover, for larger values of $\tan\beta$ the $2\mbox{ jets} + \tau$
mode dominates in all points in SUSY parameter space; see Fig.\
\ref{fig:zone35-1}.

Finally, for $\tan \beta= 35$ and $\epsilon=10$ GeV, only the channels
involving chargino to tau--majoron play a significant r\^ole, and
consequently the MSSM topologies cannot give any information.  In
other words, in this case the main contributions to the chargino mass
constraints, as seen from \fig{fig:zone35-10}, come from $\tau\tau
\ptmis$ and $2\mbox{ jets} + \tau$ topologies.  In this range of parameters,
LEP II is also able to probe chargino masses almost up to the
kinematical limit despite the presence of the irreducible $WW$
background; see section \ref{SMb}. Furthermore, for such a large value
of $\epsilon$ and smaller values of $\tan\beta$ , the chargino masses
compatible with the limits on the $\nu_\tau$ mass are not accessible
at LEP II energies, as can be seen in Fig. \ref{fig:mntau_eps}.


We summarize our results in Table \ref{t:1} where we show the 95\% CL
chargino mass limits, that can be obtained in the absence of any
signal at LEP II, for different values of the effective bilinear
$R$--parity violation parameter $\epsilon$ and two representative
values of $\tan\beta$. These bounds are the weakest constraints that
can be obtained when we vary the parameters in the ranges given by
Eqs.\ (\ref{param2}) and (\ref{param1}), and they resulted from the
analysis of each topology separately, as well as from the combined
results.  In the case where no limit was quoted in Table \ref{t:1},
the bound obtained was lower than $45$ GeV, the kinematical limit for
LEP I, although the corresponding result was used in the combined
bound.  As we can see, the {\sl combined} constraints are almost
independent of $\tan \beta$, and of the $R$--parity breaking parameter
$\epsilon$.  For small values of $\epsilon$, as expected, the chargino
mass bounds reach up to the kinematical limit, recovering exactly the
MSSM results for vanishing $\epsilon$ and $v_L$. For large $\epsilon$,
they come solely from $\tau \tau JJ$. For intermediate values,
$\epsilon \approx $ 1 GeV or so, the combination of channels is
necessary, mainly $\tau\tau \ptmis$ and $2\mbox{ jets} + \tau$
topologies.

Assuming unification of the gaugino mass parameters, we can derive
bounds on the neutralino mass from the limits on the chargino mass.
We obtained a neutralino mass limit of 38 GeV for $\tan \beta =2 $ and
48 GeV for $\tan \beta =35 $, when $\epsilon$ has the values given in
Table \ref{t:1}.


\section{Comments and Conclusions}

We studied chargino pair production and decay at LEP II
($\sqrt{s}=172$ GeV) in SUSY models with spontaneously broken
$R$--parity, characterized by the existence of the Majoron. We
performed detailed signal and background analysis in order to
determine the LEP II potential in probing physical parameters such as
chargino or neutralino masses, $m_{\chi^+}$ or $m_{\chi^0}$.  We found
that for most of the $R$--parity violating SUSY parameter space the
chargino signal can be seen up to chargino masses close to the
kinematical limit.  We explicitly verified that, as $\epsilon
 \to  0$ one recovers the MSSM chargino mass limit.  Moreover,
in analogy with standard practice, we assumed unification of the
gaugino mass parameters in order to determine the corresponding
neutralino mass limit. To improve this limit it is important to
realize that a dedicated neutralino analysis is really needed, more so
than in the corresponding MSSM case since the neutralino may exhibit
visible decays.

\begin{figure}
\begin{center}
        \mbox{\epsfig{file=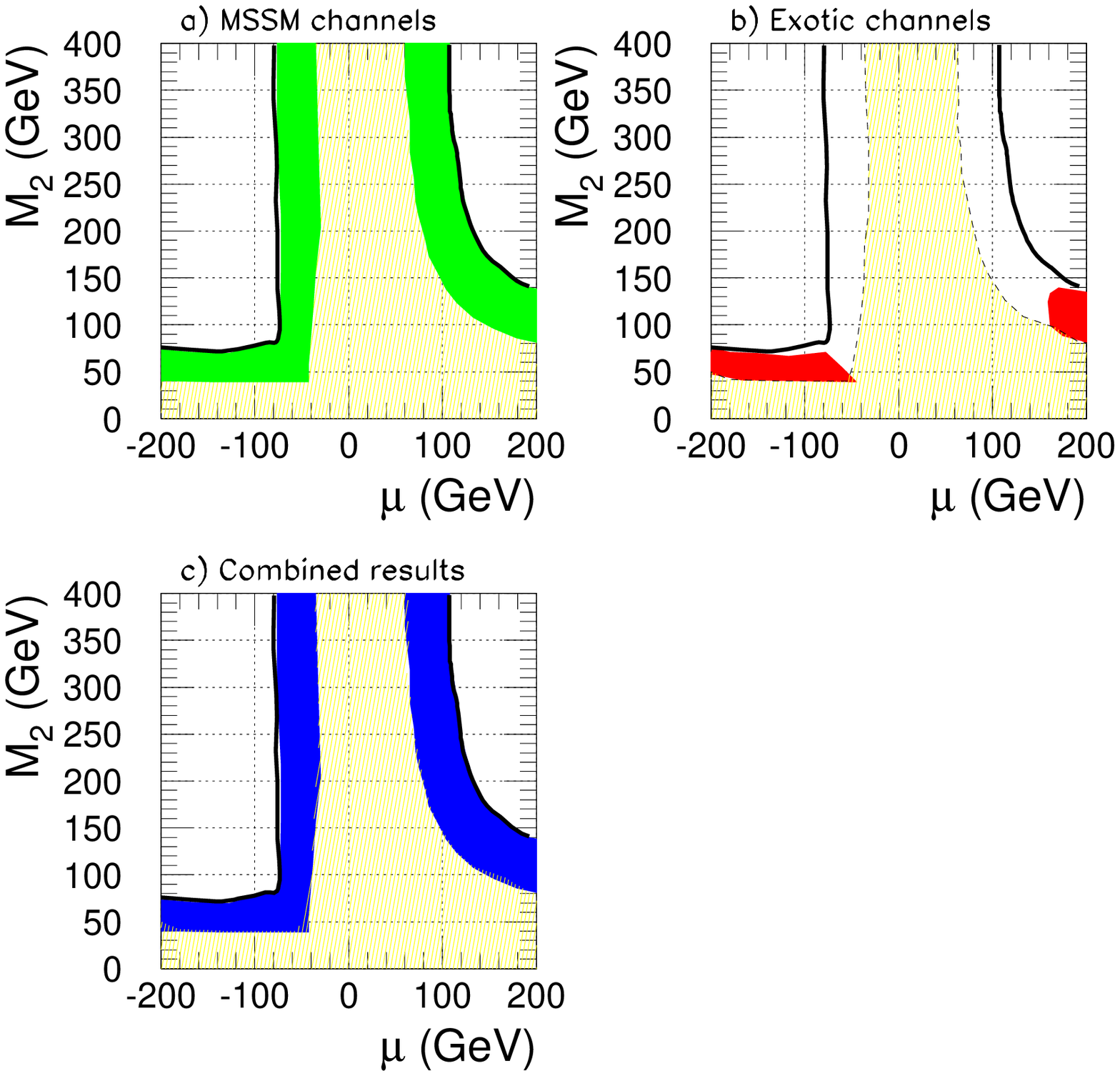,width=\linewidth}}
\end{center}
\caption{95\% CL excluded region in the ($\mu $, $M_2$) plane (dark 
  shaded areas) by the analyses of the MSSM (a) and exotic (b) channels, as
 well as the combined excluded region (c). We assumed $\tan \beta =2$,
 $\epsilon = 0.1$ GeV, $\protect\sqrt{s}=172$ GeV, and an integrated
 luminosity of 300 pb$^{-1}$.}
\label{fig:zone2-0.1}
\end{figure}

\begin{figure}
\begin{center}
        \mbox{\epsfig{file=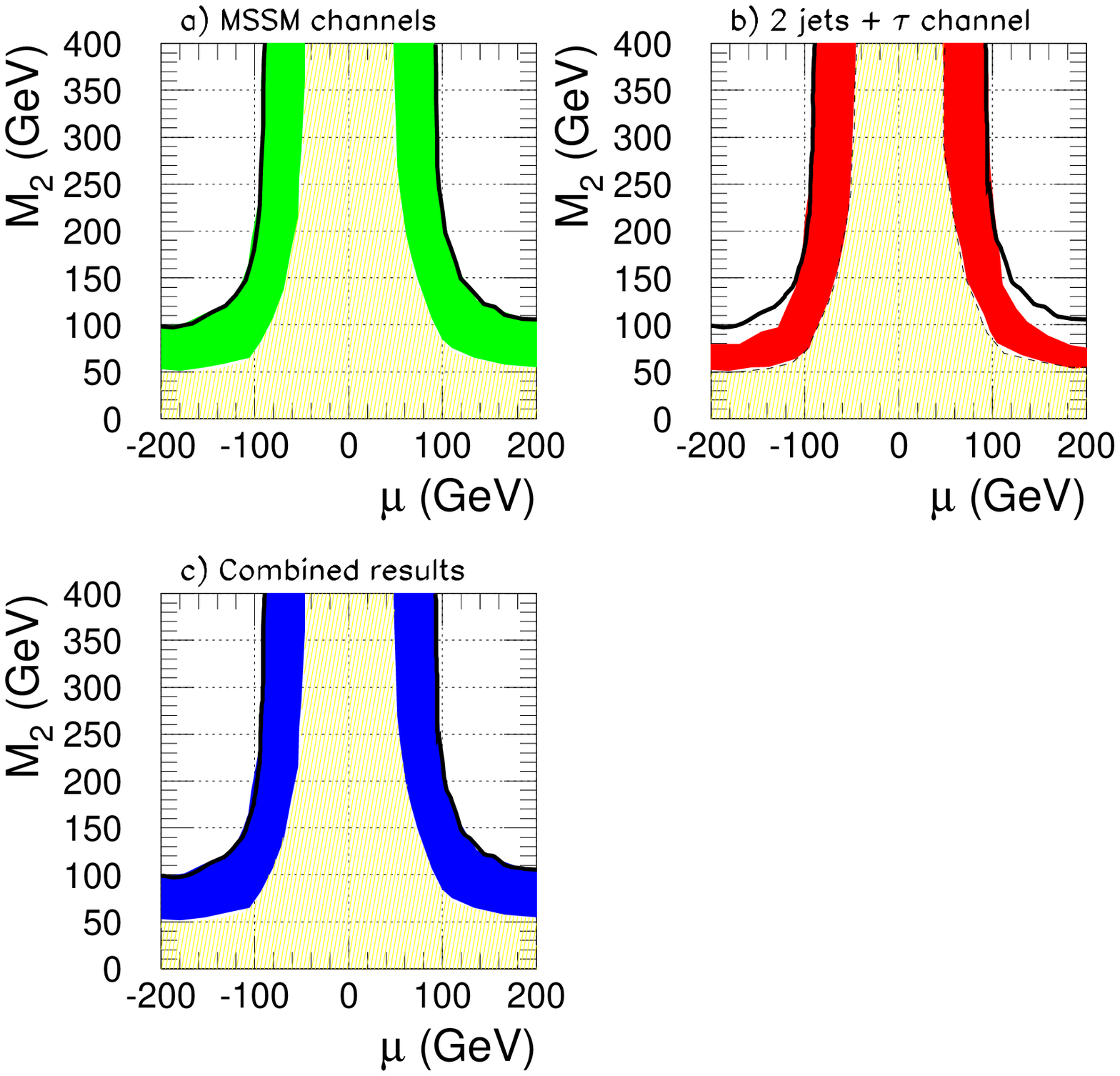,width=\linewidth}}
\end{center}
\caption{95\% CL excluded region in the ($\mu $, $M_2$) plane (dark 
  shaded areas) by the analyses of the MSSM (a) and 2 jets $+\tau$ (b)
 channels, as well as the combined excluded region (c). We assumed $\tan \beta
 =35$, $\epsilon = 0.1$ GeV, $\protect\sqrt{s}=172$ GeV, and an integrated
 luminosity of 300 pb$^{-1}$.}
\label{fig:zone35-0.1}
\end{figure}

\begin{figure}
\begin{center}
  \mbox{\epsfig{file=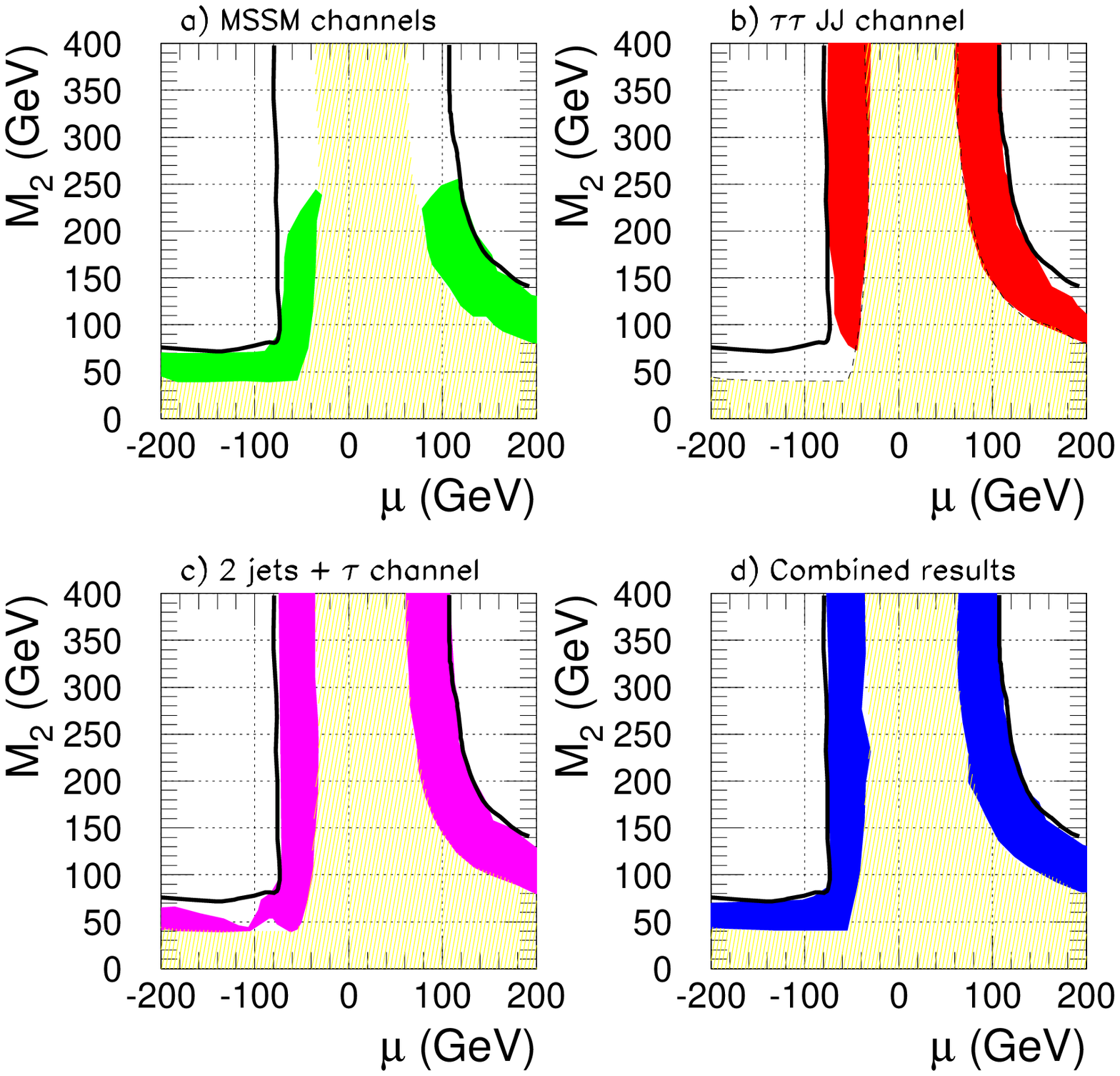,width=\linewidth}}
\end{center}
\caption{95\% CL excluded region in the ($\mu $, $M_2$) plane (dark 
  shaded areas) by the analyses of the MSSM (a), $\tau\tau JJ$ (b),
  and 2 jets $+\tau$ (c) channels, as well as the combined excluded
  region (d). We assumed $\tan \beta =2$, $\epsilon = 1$ GeV,
  $\protect\sqrt{s}=172$ GeV, and an integrated luminosity of 300
  pb$^{-1}$.}
\label{fig:zone2-1}
\end{figure}

\begin{figure}
\begin{center}
        \mbox{\epsfig{file=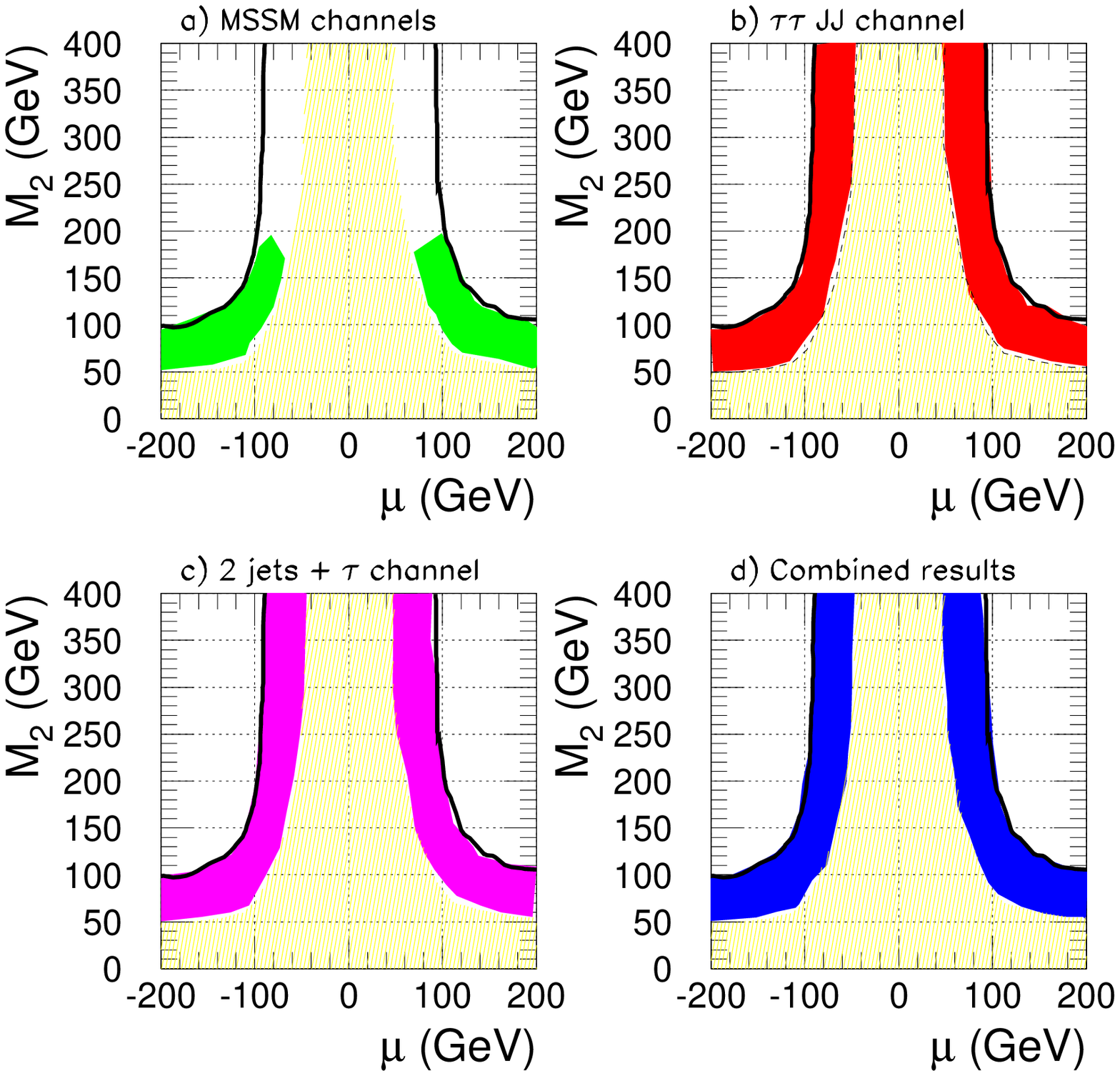,width=\linewidth}}
\end{center}
\caption{95\% CL excluded region in the ($\mu $, $M_2$) plane (dark 
  shaded areas) by the analyses of the MSSM (a), $\tau\tau JJ$ (b),
  and 2 jets $+\tau$ (c) channels, as well as the combined excluded
  region (d). We assumed $\tan \beta =35$, $\epsilon = 1$ GeV,
  $\protect\sqrt{s}=172$ GeV, and an integrated luminosity of 300
  pb$^{-1}$.}
\label{fig:zone35-1}
\end{figure}

\begin{figure}
\begin{center}
  \mbox{\epsfig{file=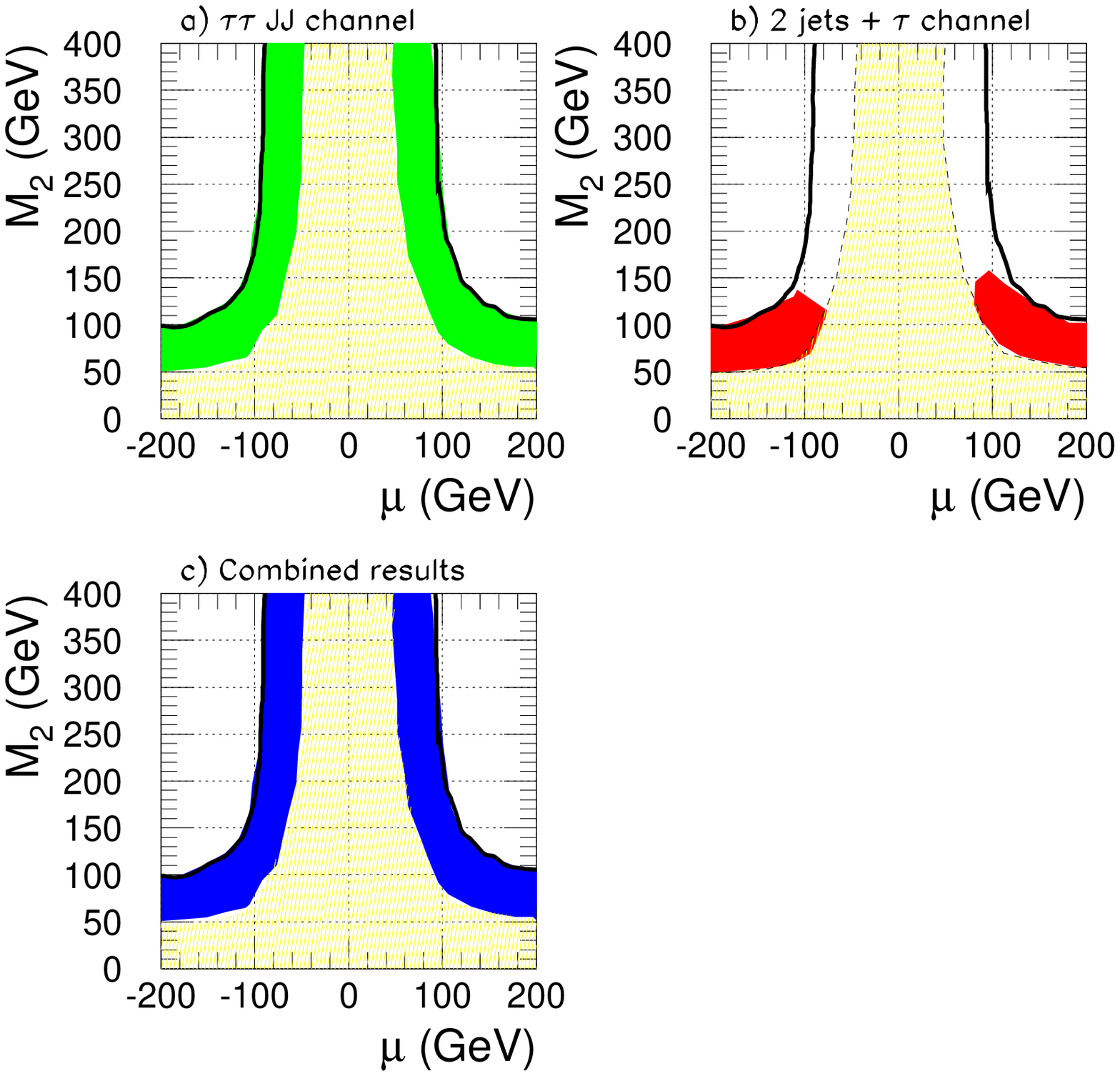,width=\linewidth}}
\end{center}
\caption{95\% CL excluded region in the ($\mu $, $M_2$) plane (dark 
  shaded areas) by the analyses of the $\tau\tau JJ$ (a) and 2 jets
  $+\tau$ (b) channels, as well as the combined excluded region
  (c). We assumed $\tan \beta =35$, $\epsilon = 10$ GeV,
  $\protect\sqrt{s}=172$ GeV, and an integrated luminosity of 300
  pb$^{-1}$.}
\label{fig:zone35-10}
\end{figure}

Our analysis show that LEP II is able to discriminate between the MSSM
and a model presenting spontaneous $R$--parity breaking in a large
region of the SUSY parameter space, if charginos are indeed observed!
For small values of $\epsilon (\simeq 0.1$ GeV) and $\tan\beta (\simeq
2)$, the exotic multi-fermion channel can be seen and therefore used
to look for $R$--parity violation when the MSSM topology is the
dominant one; see Fig.\ \ref{fig:zone2-0.1}. For larger of $\epsilon$
and $\tan\beta$, the chargino decay into $\tau J$ becomes important,
and consequently, the $2\mbox{ jets} + \tau$ and $\tau\tau\ptmis$
topologies should provide an undeniable signal for spontaneous
breaking of $R$--parity; see Figs.\ \ref{fig:zone35-0.1} to
\ref{fig:zone35-10}.

As a final remark, we have assumed in our calculations that the whole
integrated luminosity was collected at 172 GeV. Nevertheless, LEP II
has already started running at 183 GeV. This increase in energy will
enlarge the excluded area shown in our exclusion plots. However, we
leave for the experimentalists the task of doing a more detailed
analysis.


\acknowledgments

We thank M. Gandelman, J. J. Hernandez and S. Navas, for useful
discussions. We also thank T. Sjostrand for valuable discussions
and technical support in the preparation of our Monte Carlo
for the study of the $\tau \tau JJ$ signature.
 This work was supported by DGICYT under grant PB95-1077,
by the TMR network grant ERBFMRXCT960090 of the European Union, by
Conselho Nacional de Desenvolvimento Cient\'{\i}fico e Tecnol\'ogico
(CNPq), and by Fundac\~ao de Amparo \`a Pesquisa do Estado de S\~ao
Paulo (FAPESP).  M. A. G-J was supported by Spanish MEC FPI
fellowships.


\newpage

\begin{center}
\begin{table}
\begin{tabular} {||c|c|c|c|c|c|c||}
$\epsilon$ (GeV)
&
$\tan\beta$
& 
$\begin{array}{c}
$MSSM$ \\ 
$channels$
\end{array}$
& $\tau \tau +\ptmis$ & 
$\begin{array}{c}
$Exotic$ \\ 
$channels$
\end{array}$
& dijet$+\tau +\ptmis$ & 
$\begin{array}{c}
$Combined$ \\ 
$results$
\end{array}$ \\
\hline
\hline
 0  & 
$\begin{array}{c}
2 \\ 
35
\end{array}$
 & 
$\begin{array}{c}
86 \\ 
86
\end{array}$
&  
$\begin{array}{c}
- \\ 
-
\end{array}$
&  
$\begin{array}{c}
- \\ 
-
\end{array}$
&  
$\begin{array}{c}
- \\ 
-
\end{array}$&
$\begin{array}{c}
86 \\ 
86
\end{array}$  \\
\hline
 0.1  & 
$\begin{array}{c}
2 \\ 
35
\end{array}$
&
$\begin{array}{c}
84.6 \\ 
84
\end{array}$
&  
$\begin{array}{c}
- \\ 
-
\end{array}$
&  
$\begin{array}{c}
- \\ 
-
\end{array}$
&  
$\begin{array}{c}
- \\ 
61
\end{array}$
&
$\begin{array}{c}
86 \\ 
86
\end{array}$  \\
\hline
1  & 
$\begin{array}{c}
2 \\ 
35
\end{array}$
&
$\begin{array}{c}
- \\ 
-
\end{array}$
&  
$\begin{array}{c}
60 \\ 
80
\end{array}$
&  
$\begin{array}{c}
- \\ 
-
\end{array}$
&  
$\begin{array}{c}
63.7 \\ 
77
\end{array}$
&
$\begin{array}{c}
84.2 \\ 
86
\end{array}$  \\
\hline
10  & 
35
&
-
&  
86
&  
-
&  
-
&
86
  \\
\end{tabular}
\vskip 24pt
\caption{95\% CL chargino mass limits in GeV that can be derived from
negative searches at LEP II.}
\label{t:1}
\end{table}
\end{center}


\begin{references}
\bibitem{fr} 
We would like to thank F. Richard for calling our attention to this
point.
\end{references}
\end{document}